\documentclass[12pt,letterpaper]{article}
\usepackage{jheppub}
\usepackage{verbatim}
\usepackage{slashed}

\usepackage{graphicx}
\usepackage{subfigure}
\input{epsf}
\usepackage{epsfig}
\usepackage{epstopdf}
\usepackage{amsthm}
\usepackage{tikz}
\usepackage{soul}

\def\ST{{\rm ST}}
\def\sl{\text{sl}}

\def\({\left(} \def\){\right)}
\def\[{\left[} \def\]{\right]}

\def\L{\mathcal{L}}

\newcommand{\be}{\begin{equation}}
\newcommand{\ee}{\end{equation}}
\newcommand{\bea}{\begin{eqnarray}}
\newcommand{\eea}{\end{eqnarray}}
\newcommand{\ba}{\begin{eqnarray}}
\newcommand{\ea}{\end{eqnarray}}

\newcommand{\beq}{\begin{equation}}
\newcommand{\eeq}{\end{equation}}
\newcommand{\beqa}{\begin{eqnarray}}
\newcommand{\eeqa}{\end{eqnarray}}
\newcommand{\beqar}{\begin{eqnarray*}}
\newcommand{\eeqar}{\end{eqnarray*}}

\newcommand{\eg}{{\it e.g.}\ }
\newcommand{\ie}{{\it i.e.}\ }

\title{Effective $AdS_3/CFT_2$}

\author[a]{Soumangsu Chakraborty}
\author[b]{, Amit Giveon}
\author[c]{and  David Kutasov}

\affiliation[a]{Department of Physics,
Center for Cosmology and AstroParticle Physics (CCAPP)\\
The Ohio State University,
191 W Woodruff Ave, Columbus, OH 43210, USA}

\affiliation[b]{Racah Institute of Physics, The Hebrew University, Jerusalem, 91904, Israel}
\affiliation[c]{Kadanoff Center for Theoretical Physics, Enrico Fermi Institute, and Department of Physics  \\University of Chicago, 5640 S. Ellis Ave, Chicago, IL 60637, USA}

\emailAdd{soumangsuchakraborty@gmail.com}
\emailAdd{giveon@mail.huji.ac.il}
\emailAdd{dkutasov@uchicago.edu}

\abstract{In \cite{Balthazar:2021xeh}, it was pointed out that superstring theory on $AdS_3$ with $(NS,NS)$ $B$-field background and $R_{AdS}/l_s=\sqrt k<1$ is dual to a symmetric product CFT deformed by an operator in the $\mathbb{Z}_2$ twisted sector. We generalize the analysis of \cite{Balthazar:2021xeh} to $k>1$, and show that the resulting picture matches that discussed in the bosonic case in \cite{Eberhardt:2021vsx}. We argue that in the critical case, $k=1$ \cite{Giveon:2005mi}, the $\mathbb{Z}_2$ twisted deformation remains non-trivial. This resolves some confusions in the literature.
}

\begin{document}
\maketitle
 
\section{Introduction}
\label{intro}

In the 2021 paper \cite{Balthazar:2021xeh}, it was shown that superstring theory on $AdS_3$ supported by $(NS,NS)$ $B$-field flux simplifies for $R_{AdS}<l_s$, in agreement with expectations \cite{Giveon:2005mi}.
The $AdS_3/CFT_2$ duality relates it to a boundary (or spacetime) $CFT_2$ that takes a symmetric product form, with the seed of the symmetric product being the Seiberg-Witten (SW) long string theory \cite{Seiberg:1999xz}. That theory contains a non-compact direction $\phi$, which corresponds in the bulk to the radial direction in $AdS_3$. It is described by a linear dilaton CFT, with slope 
\begin{equation}\label{Qell}
Q_\ell=Q(1-k)~,\qquad Q\equiv\sqrt{2\over k}~,
\end{equation}
where $k=(R_{AdS}/l_s)^2$ is the level of the $SL(2,\mathbb{R})$ current algebra that underlies the worldsheet theory. The central charge of the linear dilaton CFT is $c_\phi=1+3Q_\ell^2$. There is also an associated fermion, $\psi_\phi$, with central charge $c_\psi=1/2$.

For $k<1$, the sign of $Q_\ell$ \eqref{Qell} is such that the string coupling for long strings, 
\be
\label{gsl}
g_\ell(\phi)\sim \exp\left(-\frac12 Q_\ell\phi\right)~,
\ee
goes to zero as $\phi\to\infty$, i.e. the long strings become free near the boundary of $AdS_3$. On the other hand, as $\phi\to-\infty$ the coupling grows, and one expects the (free) long string description to be modified. This is not surprising from the bulk point of view, since in this region the strings are not long. 

The above qualitative picture suggests that for $k<1$ the spacetime CFT should approach at large $\phi$ the symmetric product of SW CFT's, that describes the dynamics of multiple free long strings, with a deformation that grows as $\phi$ decreases. It was shown in \cite{Balthazar:2021xeh} that this is indeed the case. The form of the deformation was derived from the worldsheet theory, and shown to lead to a Liouville-type interaction, that goes to zero as $\phi\to\infty$ and grows as $\phi$ decreases. The fact that the long string picture breaks down in the region $\phi\to-\infty$ is reflected in the fact that this deformation contains an operator in the $\mathbb{Z}_2$ twisted sector of the symmetric orbifold. Thus, the spacetime CFT has in this case the property that at large positive $\phi$ it approaches a symmetric product, but at finite $\phi$ the symmetric product structure is modified, as expected from the above general considerations. For that reason, the resulting theory was referred to in \cite{Balthazar:2021xeh} as asymptotically free. 

Shortly after the appearance of \cite{Balthazar:2021xeh}, it was proposed \cite{Eberhardt:2021vsx} that the above picture can be generalized to $k>1$ (or $R_{AdS}>l_s$). In that regime, $Q_\ell$ \eqref{Qell} is negative, and therefore the long string coupling \eqref{gsl} grows as $\phi$ increases. This has been interpreted in \cite{Giveon:2005mi} as due to the fact that at large $\phi$ the spectrum is dominated by BTZ black holes, that can be thought of as strongly coupled bound states of some microscopic degrees of freedom. Thus, in this case, the description of the spacetime CFT in terms of a symmetric product of SW theories is expected to fail both at large positive and large negative $\phi$, for different reasons. 

The failure at large positive $\phi$ is not expected to be describable in terms of the symmetric product of SW CFT's, since the latter does not contain the microstates of BTZ black holes. Indeed, the BTZ entropy is strictly larger than that of the symmetric product \cite{Giveon:2005mi}. However, one may hope that it does describe well the physics at energy scales that remain finite in the limit where the spacetime central charge goes to infinity. The proposal of \cite{Eberhardt:2021vsx} was that this physics could be described by a symmetric product with a perturbation in the $\mathbb{Z}_2$ twisted sector that, like for $k<1$, modifies the region of large negative $\phi$. 

It is important to emphasize that the situation is different for $k<1$ and $k>1$. While in the former case, the deformed symmetric product of SW CFT's is believed to describe the full spacetime theory, in the latter it only describes part of that theory, namely, the states whose energies remain finite in the limit where the spacetime central charge $c\to\infty$. In that sense, one can think of it as an effective CFT. The situation is analogous to string theory in flat spacetime, where the perturbative string description captures states whose energies remain finite in the limit $g_s\to 0$, but not the black holes whose energies go like $1/g_s^2$.  

One of the consequences of the above discussion is that while the full string theory on $AdS_3$ is expected to be modular invariant in spacetime for all $k$, the deformed symmetric product cannot have this property for $k>1$, since the asymptotic density of states in the full spacetime CFT is that of black holes, which are not described by it. Therefore, while the spacetime theory proposed in \cite{Balthazar:2021xeh} for $k<1$ is modular invariant, the one of \cite{Eberhardt:2021vsx} for $k>1$ is not. 

The construction of \cite{Eberhardt:2021vsx} was done in the bosonic string, and it is natural to ask whether it can be generalized to the fermionic string, with or without spacetime supersymmetry. Some remarks on this question appeared in \cite{Eberhardt:2021vsx} and subsequent literature. The goal of this note is to study this generalization from the perspective of \cite{Balthazar:2021xeh}. The basic observation is that while that paper emphasized the case $k<1$, many aspects of its analysis are valid for all $k$, and shed new light on the case $k\ge 1$. In particular, one can identify the deformation of the SW symmetric product found in \cite{Balthazar:2021xeh} for general $k$, and use it to write down the deformed theory that describes the low-lying states (in the sense explained above) in the spacetime CFT for $k\ge 1$. 

This note is partly motivated by recent studies of the special case $k=1$, such as \cite{Giribet:2018ada,Giribet:2020mkc,Gaberdiel:2024dva}. In that case, the slope of the linear dilaton in the SW CFT \eqref{Qell} vanishes, and the leading behavior of the asymptotic density of states of black holes and fundamental strings coincide \cite{Giveon:2005mi}. 
As for $k\not=1$, it is natural to take as the starting point of the description of the spacetime CFT the SW long string theory. The remaining question is how this description is modified in the UV $(\phi\to\infty)$ and IR $(\phi\to-\infty)$ regions.

Below we argue that for $k=1$ the SW description is valid in the UV region, but is modified in the IR. In particular, we identify the $\mathbb{Z}_2$ twisted deformation that breaks the symmetric product structure at finite $\phi$. This deformation is non-normalizable, a phenomenon familiar from Liouville-type theories;  interestingly, here it appears in a theory in which asymptotically (at large $\phi$) the dilaton is constant.  

The deformation of the symmetric product CFT at finite $\phi$ is of course not surprising, since as mentioned above, it is essentially the statement that the symmetric product description fails when the strings are not long, which is the expected result for all $k$, including $k=1$.

To arrive at the above conclusions, we generalize the results of \cite{Balthazar:2021xeh} to $k\ge 1$, and those of \cite{Eberhardt:2021vsx} to the fermionic string. We start with a discussion of the general construction and then turn to some special cases. We use extensively the results and notation of \cite{Balthazar:2021xeh}, which includes many additional details, and references to earlier literature. Thus, this note is best read in conjunction with that paper.

\section{$AdS_3\times\mathcal{N}$}
\label{sec2}

To describe the fermionic string on $AdS_3$, we need to choose a unitary compact $N=1$ SCFT $\mathcal{N}$, whose central charge is determined by the criticality of the full background, 
\be
\label{cn}
\left(3+\frac6k\right)+\frac32+c_\mathcal{N}=15~.
\ee
The first two terms on the l.h.s. are the central charges of the bosonic $SL(2,\mathbb{R})$ WZW model at level $k+2$, and three free fermions $\psi^a$, $a=1,2,3$, that transform in the adjoint representation of a level $-2$ $SL(2,\mathbb{R})$ current algebra. The total level $k$ is related to the size of the $AdS$ spacetime, $R_{AdS}$, via the relation $R_{AdS}=l_s\sqrt k$.

To define the background, we need to specify a GSO projection. As reviewed in \cite{Balthazar:2021xeh}, one can construct $AdS_3$ analogs of the type 0 and type II theories in flat spacetime, by taking this projection to be (non-)chiral. In the former case, the resulting theory has BF violating tachyons, while in the latter these tachyons can be projected out from the spectrum. We are interested in the chiral case, but much of our discussion below is valid for the type 0 theory as well.

The precise form of the type II GSO projection depends on the SCFT $\mathcal{N}$. In all known examples, it acts as  $(-)^{F_L}\Omega$, where $F_L$ is the fermion number acting on the three left-moving fermions $\psi^a$ defined above, and $\Omega$ specifies the action on $\mathcal{N}$. For example, in \cite{Balthazar:2021xeh} $\mathcal{N}$ is a squashed $\mathbb{S}^3$, and $\Omega$ is given by equation (5.8) in that paper. Later, we will discuss the cases $\mathcal{N}=\mathbb{S}^3\times \mathbb{T}^4,~\mathbb{S}^3\times \mathbb{S}^3\times \mathbb{S}^1$, which contain seven additional free fermions $\psi^i$, $i=4,\cdots,10$, and $\Omega$ acts as $(-)^{F_L}$ on these fermions as well. 

Seiberg and Witten \cite{Seiberg:1999xz} showed that a long string winding the spatial circle and propagating in the bulk of $AdS_3$ is described by the non-compact CFT 
\be
\label{longs}
M_{6k}=\mathbb{R}_\phi\times \mathcal{N}~.
\ee
$\mathbb{R}_\phi$ describes the radial motion of the string, and is given by a linear dilaton CFT for a field $\phi$, with the slope \eqref{Qell}. It also includes a free fermion $\psi_\phi$, which is related to $\phi$ by an emergent supersymmetry of the SW theory. 

Using \eqref{cn}, one can check \cite{Seiberg:1999xz} that the central charge of \eqref{longs} is 
\be
\label{clong}
c(M_{6k})=6k~.
\ee
Multiple long strings propagating in the bulk are (approximately) free, and are described by the symmetric product CFT 
\be
\label{symprodcft}
\left(M_{6k}\right)^N/S_N~,
\ee
where $N$ is the number of strings. 

The CFT \eqref{longs}, \eqref{symprodcft} is singular, due to the presence of the non-compact linear dilaton factor $\mathbb{R}_\phi$. As mentioned in section \ref{intro}, the nature of the singularity is different in the two cases $k<1$ and $k>1$. In the former case, the symmetric product description becomes better and better as $\phi\to\infty$, since the coupling between long strings, \eqref{gsl}, goes to zero there. In the region $\phi\to-\infty$ we expect the SW description to break down, since a wound string near the center of $AdS_3$ is not long. 

For $k>1$, the description \eqref{symprodcft} is expected to break down as $\phi\to-\infty$, for the same reason as above, but it is also expected to break down as $\phi\to\infty$. Indeed, as discussed in section \ref{intro}, in this case, the long string coupling \eqref{gsl} grows in this region, and the correct description there is in terms of microstates of BTZ black holes. Thus, for $k>1$ one expects the symmetric product description to hold in a finite interval in $\phi$, with modifications at large $|\phi|$. The modification at large negative $\phi$ is expected to be the same as for $k<1$, while that at large positive $\phi$ should only be present for this case.

\subsection{$k<1$}
\label{sec21}

To generalize the analysis of \cite{Balthazar:2021xeh} to $k>1$ we next briefly review the relevant part of that analysis for $k<1$. The starting point is the operator\footnote{As is standard for asymptotically linear dilaton theories, \eqref{ebphi} is the asymptotic behavior of the operator for large positive $\phi$. It defines the operator uniquely, but its precise form for finite $\phi$ is different. In particular, for real $p$ it contains both incoming and outgoing components.}
\be
\label{ebphi}
e^{\beta\phi}
\ee
in the SW theory \eqref{longs}. Here $\beta$ is given by 
\be
\label{defbeta}
\beta=-\frac{Q_\ell}{2}+ip~,
\ee
where the linear dilaton slope $Q_\ell$ \eqref{Qell} is positive for $k<1$. The radial momentum $p$ is real for delta-function normalizable operators. It can be taken to be positive, since incoming and outgoing states are not independent.

The operator \eqref{ebphi} can be written as 
\be
\label{SWpsi}
e^{\beta\phi}=g_\ell(\phi)\Psi(\phi)~,
\ee
where $g_\ell(\phi)$ is given by \eqref{gsl}, and $\Psi(\phi)$ is the wavefunction, 
\be
\label{Psibeta}
\Psi(\phi)=e^{\left(\beta+\frac{Q_\ell}2\right)\phi}=e^{ip\phi}
\ee
(see eq. (4.15) in \cite{Balthazar:2021xeh}). Non-normalizable operators can be obtained by continuation to real positive $ip$. 

The dimension of the operator \eqref{ebphi} in the spacetime CFT is 
\be
\label{dimexp}
h_{\ST}=\bar h_{\ST} = -\frac12\beta(\beta+Q_\ell)= \frac{p^2}{2}+\frac{Q_\ell^2}{8}~.
\ee
The $(-1,-1)$ picture worldsheet vertex operator corresponding to \eqref{ebphi} is given by equation (4.9) in \cite{Balthazar:2021xeh},
\be
\label{ebphiws}
e^{\beta\phi}~~\longleftrightarrow~~e^{-\varphi-\bar\varphi} \, e^{i(H_\sl+\bar H_\sl)} \, \Phi_{j;m,\bar m}^{(-1)}~.
\ee
The quantities on the r.h.s. of \eqref{ebphiws} are defined and discussed in \cite{Balthazar:2021xeh}; we will not review them here. The mass-shell condition of the bulk operator on the r.h.s. is 
\be
\label{massshell}
m=\bar m=\frac{j(j-1)}{k}+\frac{k+2}{4}~.
\ee
The spacetime scaling dimension of this operator is
\be
\label{stscale}
h_{\ST}=\bar h_{\ST}=\frac{k}{2}-m=-\frac{j(j-1)}{k}+\frac{k-2}{4}~.
\ee
Delta-function normalizable operators correspond to 
\be
\label{longsp}
j=\frac12+is~. 
\ee
The vertex operator on the r.h.s. of \eqref{ebphiws} describes a long string carrying radial momentum proportional to $s$, which can be taken to be positive since the incoming and outgoing states are related by reflection. 

Plugging \eqref{longsp} into \eqref{stscale}, we see that the bulk operator \eqref{ebphiws} carries spacetime scaling dimension
\be
\label{hhlong}
h_{\ST}=\bar h_{\ST}=\frac{s^2}{k}+\frac{Q_\ell^2}{8}~.
\ee
Comparing \eqref{dimexp} with \eqref{hhlong}, we see that they agree if we take
\be
\label{matchwave}
p=s\sqrt\frac{2}{k}=sQ~.
\ee
As discussed in \cite{Balthazar:2021xeh}, this relation can be interpreted as the statement that the bulk and boundary wavefunctions have the same radial dependence, see equations (4.15), (4.16). This resolves the sign ambiguity in going from \eqref{dimexp}, \eqref{hhlong} to \eqref{matchwave}.

In the SW theory \eqref{longs}, the operator $\partial_x\phi$ is holomorphic, $\partial_{\bar{x}}\partial_x\phi=0$. To probe whether in string theory on $AdS_3\times\mathcal{N}$ the symmetric product \eqref{symprodcft} is deformed, we compute this quantity directly in the bulk description. This calculation was done in \cite{Balthazar:2021xeh}, with the result given in eq. (7.5),
\begin{equation}\label{ddphimap}
\partial_{\bar{x}}\partial_x\phi 
~\longleftrightarrow~
    e^{-\varphi-\bar\varphi}\,
e^{i(H_{sl}+\bar H_{sl})}\, 
(\partial\varphi+i\partial H_{sl}) 
(\bar\partial\bar\varphi+i\bar\partial \bar{H}_{sl})\,
\Phi^{(-1)}_{1-\frac{k}{2};\frac{k}{2}-1,\frac{k}{2}-1}~.
\end{equation}
It is useful for future reference to recall some properties of \eqref{ddphimap}. The operator on the r.h.s. of this equation has $j=1-\frac k2$, which for $k<1$ satisfies $j>\frac12$. For this value of $j$, the non-normalizable worldsheet wavefunction behaves like 
\be
\label{psinon}
\Psi_{\rm non-norm}(\phi)\simeq e^{\frac Q2(1-k)\phi}
\ee
for large positive $\phi$ (see eq. (4.16) in \cite{Balthazar:2021xeh}). However, in \eqref{ddphimap} one also has $m=\bar m=-j$ and therefore this operator sits on an LSZ pole \cite{Aharony:2004xn}. As discussed in \cite{Balthazar:2021xeh}, what really appears on the r.h.s. of \eqref{ddphimap} is the residue of this pole, i.e. the corresponding normalizable operator, which behaves at large $\phi$ like 
\be
\label{psinor}
\Psi_{\rm norm}(\phi)\simeq e^{-\frac Q2(1-k)\phi}~.
\ee
Note that since $k>\frac12$ (see \eqref{cn}), $j$ in \eqref{ddphimap} also satisfies the constraint $j<(k+1)/2$ (eq. (2.48) in \cite{Balthazar:2021xeh}), which is necessary for the state to be normalizable.

An important part of the analysis of \cite{Balthazar:2021xeh} is the observation that the normalizable operator with $w=-1$ \eqref{ddphimap} has an FZZ dual with $w=-2$ (see the discussion around equations (2.62) and (7.10) in \cite{Balthazar:2021xeh}),
\begin{equation}\label{ddphifzz}
    e^{-\varphi-\bar\varphi}\,
e^{i(H_{sl}+\bar H_{sl})}\, 
(\partial\varphi+i\partial H_{sl}) 
(\bar\partial\bar\varphi+i\bar\partial \bar{H}_{sl})\,
\Phi^{(-2)}_{k;k,k}~.
\end{equation}
This operator has $j=k$, which is also in the range $\frac12<j<(k+1)/2$ for $\frac12<k<1$. The corresponding normalizable wavefunction goes like 
\be
\label{psinor2}
\Psi_{\rm norm}(\phi)\simeq e^{-Q(k-\frac12)\phi}~.
\ee
As usual, FZZ duality means that the normalizable operator contains both contributions \eqref{ddphimap} and \eqref{ddphifzz}. It is normalizable since both of the wavefunctions, \eqref{psinor} and \eqref{psinor2}, decay exponentially as $\phi\to\infty$. 

To identify the operators \eqref{ddphimap}, \eqref{ddphifzz} in the spacetime theory \eqref{symprodcft}, we use the observation in \cite{Balthazar:2021xeh}, that the worldsheet and spacetime wavefunctions coincide. The operator \eqref{ddphimap} has winding $-1$ and therefore it lives in the seed of the symmetric product, the  SW theory \eqref{longs}. Its $\phi$ dependence can be computed by multiplying the wavefunction \eqref{psinor} by $g_\ell(\phi)$ \eqref{gsl}, which means that it goes like $\exp(-Q_\ell\phi)$. It was argued in \cite{Balthazar:2021xeh} that it corresponds in the SW theory to the operator 
\be
\label{qqll}
\partial_x\phi\partial_{\bar{x}}\phi e^{-Q_\ell\phi}
\ee
(see eq. (7.6) in \cite{Balthazar:2021xeh}). 

One way to think about \eqref{qqll} is to note that the operator $\exp(-Q_\ell\phi)$ has spacetime scaling dimension $(0,0)$ (see \eqref{dimexp}), while the operator on the r.h.s. of \eqref{ddphimap} is a spacetime Virasoro primary with scaling dimension $(1,1)$. Furthermore, this operator is manifestly insensitive to the particular theory $\mathcal{N}$ in \eqref{longs}. The operator \eqref{qqll} is the unique candidate with all these properties. 

Thus, following \cite{Balthazar:2021xeh} we conclude that the worldsheet calculation leads to a deformation of the SW Lagrangian by a term that modifies the metric for $\phi$,
\be
\label{deformblock}
\L=\L_0+\lambda \partial_x\phi\partial_{\bar{x}}\phi e^{-Q_\ell\phi}~.
\ee
This deformation leads to a modified equation of motion for $\phi$, whose leading behavior for large $\phi$ is
\be
\label{ddphi}
\partial_{\bar{x}}\partial_x\phi\simeq \partial_x\phi \partial_{\bar{x}}\phi e^{-Q_\ell\phi}~,
\ee
which is precisely what we found in our worldsheet analysis, \eqref{ddphimap}. The role of the deformation \eqref{deformblock} is discussed further in \cite{Balthazar:2021xeh} (see eq. (7.7) and the subsequent discussion). An important feature for our properties is that it acts within the seed of the SW orbifold \eqref{symprodcft}, modifying it at finite $\phi$.

The operator \eqref{ddphifzz} has $w=-2$ and thus lives in the $\mathbb{Z}_2$ twisted sector of the orbifold \eqref{symprodcft}. Using the results of \cite{Balthazar:2021xeh},
one can compute the radial profile of this operator as follows. First, in the product of two copies of the SW theory \eqref{longs}, one can define two fields $\phi_1$, $\phi_2$, which are the $\phi$ fields of the two copies. The field 
\be
\label{phiave}
\phi_{\rm ave}=\frac12(\phi_1+\phi_2)
\ee
is invariant under the $\mathbb{Z}_2$ exchanging the two copies, and therefore is in the untwisted sector of the orbifold.\footnote{This field is defined (for general $w$) in eq. (4.58) of \cite{Balthazar:2021xeh}. For $w=2$ it is convenient to express it in terms of the field  $\phi_S=\sqrt2\phi_{\rm ave}$ defined in eq. (7.16) of that paper. As explained there, this field is canonically normalized, and has linear-dilaton slope $\sqrt2 Q_\ell$.} This field is identified in subsection 4.3 of  \cite{Balthazar:2021xeh} with the worldsheet field $\phi$. Using this identification, the $\phi$ dependence of the boundary CFT operator corresponding to \eqref{ddphifzz} is obtained in appendix~C of \cite{Balthazar:2021xeh}. One finds that the result of that calculation can be described by multiplying the wavefunction \eqref{psinor2} by $g_\ell^2$ (one power of $g_{\ell}$, \eqref{gsl}, for each copy of $M_{6k})$. The resulting profile of the $\mathbb{Z}_2$ twisted operator goes like 
\be
\label{profztwo}
e^{-\frac{1}{\sqrt{2k}}\phi_{\rm ave}}~.
\ee
As a check, this result agrees with eq. (7.19) in \cite{Balthazar:2021xeh}, which describes this operator for the particular class of theories studied in that paper. 

To describe the $\mathbb{Z}_2$ twisted operator more precisely, it is useful to follow the approach used in \cite{Eberhardt:2021vsx} for the bosonic case. Instead of constructing the twisted operator, one can construct the corresponding untwisted sector operator in the covering space. The dimensions of the two are related by the standard formula \cite{Klemm:1990df},
\be
\label{h1w}
h_w=\frac1w h_1+\frac{c}{24}\left(w-\frac1w\right)~,
\ee
which in our case ($w=2$, $c=6k$) takes the form 
\be
\label{h12}
h_2=\frac12 h_1+\frac{3k}{8}~.
\ee
Using the results in appendix C of \cite{Balthazar:2021xeh}, one can show that the zero-mode part of that operator goes like
\be
\label{profz1}
e^{-\sqrt\frac k2\phi}~.
\ee
Its scaling dimension is $(h_1,h_1)$, with
\be
\label{honephi}
h_1=-\frac12\beta(\beta+Q_\ell)=\frac12-\frac{3k}4~,
\ee
where in the second equality we used \eqref{profz1}, $\beta=-\sqrt\frac k2$. The full $\mathbb{Z}_2$ twisted operator \eqref{ddphifzz} has dimension $(1,1)$, thus we see from \eqref{h12}, \eqref{honephi}, that in the covering space its dual is given by \eqref{profz1} multiplied by an operator of dimension $\frac32$. The remaining task is to find that operator. 

To do that, one notes that the corresponding worldsheet operator, \eqref{ddphifzz}, only depends on the $AdS_3$ part of the background, and not on the detailed structure of the compact CFT $\mathcal{N}$ \eqref{longs}. The analysis of \cite{Argurio:2000tb} then shows that the corresponding untwisted sector operator doesn't depend on $\cal N$ as well (see also (2.63)--(2.65) and (4.52) in \cite{Balthazar:2021xeh}). Thus, we expect the dimension $\frac32$ operator in the SW CFT mentioned above to only depend on the $\mathbb{R}_\phi$ factor in \eqref{longs}, and the full operator to be a linear combination of 
\be
\label{profz1full}
(\partial_x\psi_\phi) e^{-\sqrt\frac k2\phi}\;\;\;{\rm and} \;\;\;\psi_\phi(\partial_x\phi) e^{-\sqrt\frac k2\phi}~.
\ee
We will not try to determine the precise linear combination here.

\subsection{$k\ge1$}

In this subsection, we will discuss the generalization of the analysis of \cite{Balthazar:2021xeh}, summarized in subsection \ref{sec21}, to the case $k>1$, and comment on the transition between the two regimes at  $k=1$. An important feature of this discussion is the sign flip of $Q_\ell$ \eqref{Qell} at $k=1$. This means that for $k>1$ the long string coupling $g_\ell(\phi)$ \eqref{gsl} grows as $\phi\to\infty$, i.e. as we approach the boundary of $AdS_3$. Our main goal is to discuss how different elements of the discussion in section \ref{sec21} generalize to this case. 

The operator map \eqref{ebphiws} -- \eqref{matchwave} is expected to be valid for all $k>\frac12$. Therefore, so does the map \eqref{ddphimap} that follows from it. However, in this case, the value of $j$ in this equation, $j=1-\frac k2$, lies outside the unitarity band, and therefore it becomes non-normalizable. This is readily seen from the form of the wavefunction \eqref{psinor}, which diverges exponentially as $\phi\to\infty$.  Of course, in this case, this region is strongly coupled, \eqref{gsl}, so it's not clear whether this notion of normalizability is sensible. 

In the dual boundary theory, one expects, as in section \ref{sec21}, the Lagrangian of the seed of the symmetric product to be deformed by the term \eqref{deformblock}. Since $Q_\ell$ is now negative, this term grows at large $\phi$. As mentioned in \cite{Balthazar:2021xeh}, one can think of \eqref{deformblock} as a deformation of the metric for $\phi$ (see the discussion around eq. (7.9) in \cite{Balthazar:2021xeh}). One can absorb it into a redefinition of $\phi$, and therefore it is not clear whether it has any physical significance. It is intriguing that it grows in the region where the SW theory is expected to break down due to the existence of black holes, but the significance of this observation remains to be understood. 

As explained in section \ref{sec21}, the deformation \eqref{ddphimap} is accompanied by its FZZ dual, \eqref{ddphifzz}. This deformation corresponds in the boundary theory to a $\mathbb{Z}_2$ twisted sector operator that goes like 
\eqref{profztwo} at large $\phi_{\rm ave}$. It goes exponentially to zero at large $\phi_{\rm ave}$, and therefore does not disrupt the symmetric product structure at large $\phi$. On the other hand, as $\phi_{\rm ave}$ decreases, the $\mathbb{Z}_2$ twisted deformation becomes more important. Thus, the symmetric product structure breaks down for finite $\phi_{\rm ave}$. This effect does not distinguish between $k<1$ and $k>1$, since it depends smoothly on $k$,  in agreement with the qualitative picture presented in section \ref{intro}. 

It is interesting to consider the special case $k=1$, in which $Q_\ell$ \eqref{Qell} vanishes. One consequence of this is that the leading correction of the metric for $\phi$ in the SW theory, \eqref{deformblock}, is independent of $\phi$, and thus at least to this order, this theory is invariant under translations of $\phi$. We expect this translation invariance of the seed CFT to be violated by higher-order contributions to the metric of $\phi$, but the analysis of \cite{Seiberg:1999xz} suggests that these contributions go to zero at large $\phi$. The $\mathbb{Z}_2$ twisted deformation with the profile \eqref{profztwo}, or the corresponding deformation \eqref{profz1full} in the covering space, also goes to zero at large $\phi$ and breaks translation invariance at finite $\phi$.

Note that there is a potential subtlety in the above discussion. For $k\not=1$, the coefficient of the above deformation is known to be non-zero, \cite{Balthazar:2021xeh,Eberhardt:2021vsx}, but one may wonder whether this coefficient vanishes for $k=1$. In principle, one can settle this issue by a calculation of the two-point function of the operator on the r.h.s. of \eqref{ddphimap}, \eqref{ddphifzz}. This calculation depends on various normalizations in the definition of these operators and the path integral, and is thus subtle. However, one can argue indirectly that it is non-zero, as follows. 

If the operator on the r.h.s. of \eqref{ddphimap} vanished for $k=1$, the SW theory would have symmetries that the worldsheet analysis of this theory doesn't have. For example, it would have a holomorphic conserved current $\partial_x\phi$, and an $N=1$ superconformal symmetry that relates $\psi_\phi$ and $\phi$. It would also mean that the symmetric product structure is not violated for any $\phi$, which is in contradiction with the worldsheet analysis. 

Indeed, in string theory on $AdS_3\times\mathcal{N}$ there are in general normalizable bound states, that in the construction of \cite{Balthazar:2021xeh,Eberhardt:2021vsx} live near the Liouville-type wall that affects the region $\phi\to-\infty$. These states are known to violate the symmetric product structure of the spectrum. For example, they do not satisfy relations like \eqref{h1w}, \eqref{h12}. In the construction of \cite{Balthazar:2021xeh,Eberhardt:2021vsx} this is very natural, since the wall, which is responsible for their existence, violates the symmetric product structure, but if this wall is absent, it would be very hard to understand the origin and pattern of energies of these states. 

Thus, we conclude that for $k=1$ the spacetime CFT corresponding to string theory on $AdS_3\times\mathcal{N}$ is a symmetric product of SW theories, \eqref{longs}, \eqref{symprodcft}, with a deformation that breaks the symmetric product structure and $\phi$ translation invariance, just like for $k\not=1$. We have constructed the deforming operator in the worldsheet language, as well as in the symmetric orbifold.

\section{Examples}
\label{sec3}

In this section, we will demonstrate the discussion of section \ref{sec2} in a few classes of models that have been studied in the literature. 

\subsection{$AdS_3\times \mathbb{S}^3\times \mathbb{T}^4$}
\label{sec31}

In this case, the level $k$ is an integer larger than one, and the SW theory \eqref{longs} is
\be
\label{334}
M_{6k}=\mathbb{R}_\phi\times \mathbb{S}^3\times \mathbb{T}^4~.
\ee
In addition to a bosonic sigma model on \eqref{334}, the SW theory contains eight left and right-moving free fermions, that are related to the bosons by $N=4$ supersymmetry. The SUSY generators for this model are described in \cite{Seiberg:1999xz}. 

The marginal deformation of the symmetric product CFT \eqref{symprodcft} is in this case the top component of a superfield, whose bottom component has spacetime scaling dimension $(\frac12,\frac12)$. One can describe this operator in two different ways: as an operator in the SW CFT \eqref{334} in covering space, or as an operator in the $\mathbb{Z}_2$ twisted sector. In the former description, this operator takes the form 
\be
\label{lowcomp}
S_\alpha \bar S_{\bar\alpha} e^{-\sqrt\frac k2\phi}~,
\ee
where $S_\alpha$ is a spin field for the eight free left-moving fermions, and similarly for $\bar S_{\bar\alpha}$. $S_\alpha$ is an eight component spinor. Four of its components transform as two doublets of the $SU(2)_L$ R-symmetry, which are the operators of interest for this construction \cite{Eberhardt:2021vsx}. As a check, the scaling dimension of \eqref{lowcomp} is (see \eqref{honephi})
\be
\label{honelower}
h_1=1-\frac{3k}4~,
\ee
and plugging into \eqref{h12} gives $h_2=1/2$, as appropriate for the bottom component of a superfield whose top component is a modulus. 

As mentioned above, we can also construct the desired operator directly in the $\mathbb{Z}_2$ twisted sector. In that description, it takes the form 
\be
\label{Z2mod}
\sigma_{\phi_A}\sigma_{WZW}\sigma_{\mathbb{T}^4}s_\alpha\bar s_{\bar\alpha}e^{-\frac1{2\sqrt k}\phi_S}~.
\ee
Here $\sigma_{\phi_A}$ is the twist field for $\phi$, described in section 7 of \cite{Balthazar:2021xeh}, which has (left and right) scaling dimension $1/16$, $\sigma_{WZW}$ is the analogous twist field for $SU(2)_{k-2}$ WZW, whose scaling dimension is $3(k-2)/16 k$, and similarly for $\sigma_{\mathbb{T}^4}$, which has dimension $1/4$. $s_\alpha$ is the twist field for the eight SW fermions; it has scaling dimension $1/2$. Finally, the exponential of $\phi_S$ is as in \eqref{profztwo} (and footnote 2). Its scaling dimension is given by $\frac3{8k}-\frac12$. Adding up all the contributions, we find that the scaling dimension of \eqref{Z2mod} is $1/2$, as expected. 

\medskip

\noindent
Note that 
\begin{itemize}
\item The operator \eqref{Z2mod} is the analog of the one in eq. (7.19) in \cite{Balthazar:2021xeh}, for the class of models described in this subsection. 
\item Acting on \eqref{Z2mod} once with the spacetime supercharges on the left and the right gives the modulus in the $\mathbb{Z}_2$ twisted sector that produces the wall that shields the region $\phi\to-\infty$,  discussed above. 
\item The operators \eqref{lowcomp} and \eqref{Z2mod} are related by the covering space construction. 
\item In the bosonic case, it was argued in \cite{Eberhardt:2021vsx} that normalizable states owe their existence to the deformation of the SW CFT (the wall), and their wavefunctions are localized near the wall. We expect the evidence for this claim presented in \cite{Eberhardt:2021vsx} to have a counterpart in the supersymmetric case. In both cases, the fact that the pattern of normalizable states does not have a symmetric product structure is understood as due to the fact that the wall involves twisted sector operators.
\end{itemize}

\subsection{$AdS_3\times \mathbb{S}^3\times \mathbb{S}^3\times\mathbb{S}^1$}
\label{sec32}

String theory on $AdS_3\times \mathbb{S}^3\times \mathbb{S}^3\times\mathbb{S}^1$ has been studied by a number of groups, who made various suggestions for the dual CFT (see e.g. \cite{Elitzur:1998mm,deBoer:1999gea,Gukov:2004ym,Tong:2014yna,Eberhardt:2017fsi,Eberhardt:2017pty,Witten:2024yod}). This theory is interesting for a number of reasons. One is that it has a large $N=4$ superconformal symmetry, in contrast to the theory of subsection \ref{sec31} that has a small $N=4$ symmetry. Another is that in this theory one can have $k=1$. Indeed,  $k$ is given in this case by 
\be
\label{formk}
\frac 1k=\frac1k_1+\frac1k_2~,
\ee
where $k_1$ and $k_2$ are the levels of the $SU(2)$ current algebras on the two three-spheres in the background. Since $k_1, k_2\ge 2$, $k$ \eqref{formk} is in general larger than one, and is equal to one in the minimal case $k_1=k_2=2$. 

Much of the analysis of subsection \ref{sec31} goes through in this case. The SW theory is 
\be
\label{3331}
M_{6k}=\mathbb{R}_\phi\times \mathbb{S}^3\times \mathbb{S}^3\times\mathbb{S}^1~,
\ee
and the analog of \eqref{lowcomp} takes the same form as there, with $S_\alpha$ the spin field for the eight free fermions in \eqref{3331}. The analog of \eqref{Z2mod} takes in this case the form  
\be
\label{Z2modnew}
\sigma_{\phi_A}\sigma_{WZW}\sigma_{\mathbb{S}^1}s_\alpha\bar s_{\bar\alpha}e^{-\frac1{2\sqrt k}\phi_S}~,
\ee
where $\sigma_{WZW}$ is now the twist field for the bosonic $SU(2)_{k_1-2}\times SU(2)_{k_2-2}$, whose scaling dimension is $3(k_1-2)/16 k_1+3(k_2-2)/16 k_2$. Adding up all the dimensions in \eqref{Z2modnew} again gives the correct answer, $h=1/2$. 

Thus, we conclude that for all (integer) values of $k_1, k_2\ge2$, the SW orbifold \eqref{symprodcft}, \eqref{3331}, is deformed as discussed above, with the  $\mathbb{Z}_2$ twisted sector deformation obtained by acting with one left-moving and one right-moving supercharge on \eqref{Z2modnew}. In particular, this is the case for $k=1$, i.e. $k_1=k_2=2$ \eqref{formk}. 

Some previous studies suggested the possibility that for $k=1$ the SW orbifold remains unperturbed \cite{Giribet:2018ada,Gaberdiel:2024dva}. As explained above, there is no reason to expect this to be the case. The candidate deformation \eqref{Z2modnew} (or, more precisely, its top component) has a smooth continuation to $k=1$, and setting its coefficient to zero would be incompatible with the symmetry structure and spectrum of normalizable states in this model. Let us briefly comment on the latter. 

In \cite{Argurio:2000tb}, it was shown that this model has normalizable states with $w=0$ (or equivalently, by FZZ duality, $w=1$) for any $k$; see the discussion around eq. (4.21)\footnote{For $k=1$, these states are in the Ramond sector on the worldsheet.}. The authors of \cite{Argurio:2000tb} focused on BPS states, but it is straightforward to see that there are non-BPS states as well. In particular, by turning on momentum on the $\mathbb{S}^1$, one can construct normalizable non-BPS states with any scaling dimension in some finite range. The resulting states have analogs in higher winding sectors, but these analogs do not follow the symmetric product pattern \eqref{h1w}. Understanding these states requires the deformation we constructed, and would not be possible without it.

\subsection{General $N=2$ superconformal backgrounds}
\label{sec33}

As is well known \cite{Giveon:1999jg,Berenstein:1999gj,Giveon:2003ku}, a large class of $AdS_3\times\mathcal{N}$ backgrounds that give rise to $(2,2)$ superconformal spacetime CFT's can be constructed by starting with string theory on  
\begin{equation}\label{ma}
AdS_3\times\mathbb{S}^1\times{\cal M}~,
\end{equation}
with ${\cal M}$ a $(2,2)$ worldsheet CFT. The existence of spacetime supercharges leads to a chiral GSO projection, that acts as an orbifold by a discrete group $\Gamma$ on the background \eqref{ma}. Thus, the compact CFT $\mathcal{N}$ takes the form 
$\mathcal{N}=(\mathbb{S}^1\times \mathcal{M})/\Gamma$, and the SW theory is \eqref{longs}. 

The authors of \cite{Balthazar:2021xeh} considered the case where ${\cal M}=SU(2)/U(1)$, an $N=2$ minimal model, and constructed the analog of \eqref{Z2mod}, \eqref{Z2modnew} for this case. It is given by eq. (7.19) in that paper. For a general $N=2$ SCFT $\cal M$ one has 
\begin{equation}\label{sigmaa}
\Sigma^\pm=\exp\left[-{1\over 2\sqrt{k}}(\phi_S\mp iY_S)\right](\sigma_{\phi_A}\sigma_{Y_A}\sigma^\pm_{\psi_{A}})\Sigma^\pm_{\cal M}~,
\end{equation}
where the only new element relative to \cite{Balthazar:2021xeh} is $\Sigma^\pm_{\cal M}$, the (anti-)chiral $\mathbb{Z}_2$ twist field that creates the BPS twisted ground states in ${\cal M}^2/\mathbb{Z}_2$.  

In \eqref{sigmaa}, we exhibited explicitly only the left-moving contribution. Adding the right-movers
we find four dimension $\left({1\over 2},{1\over 2}\right)$ operators, $\Sigma^{\pm,\pm}$, that belong to the four sectors (c,c), (c,a), (a,c), (a,a). To construct the modulus \eqref{ddphifzz}, we need to act with the spacetime supercharges and take the appropriate combination, as was done in equations (7.12) and (7.13) of \cite{Balthazar:2021xeh}.

In the examples of subsections \ref{sec31} and \ref{sec32}, ${\cal M}={SU(2)\over U(1)}\times\mathbb{T}^4$ and ${SU(2)\times SU(2)\over U(1)}\times\mathbb{S}^1$, respectively. 
One can check that in these cases \eqref{sigmaa} is equivalent to \eqref{Z2mod} and \eqref{Z2modnew}, respectively. Another interesting class of theories corresponds to ${\cal M}=\mathbb{T}^2$, which gives rise to a family of theories with $k=1$. At a particular point in the Narain moduli space of the $\mathbb{T}^2$ theory, the supersymmetry is enlarged from $N=2$ to large $N=4$, and one finds the theory with $k_1=k_2=2$ discussed in the previous subsection. In this class of theories, the deformation of the symmetric product $(M_6)^N/S_N$ is necessary for similar reasons. For example, \eqref{sigmaa} breaks the $(4,4)$ supersymmetry of an undeformed 
$(\mathbb{R}\times\mathbb{T}^3)^N/S_N$ SCFT to $(2,2)$.

\section{Discussion}

In \cite{Balthazar:2021xeh}, it was shown that in string theory on $AdS_3$ with $(NS,NS)$ $B$-field, there is a strong case for the hypothesis that in backgrounds with $R_{AdS}<l_s$ (or $k<1$), the spacetime $CFT_2$ is given by the Seiberg-Witten orbifold \eqref{longs}, \eqref{symprodcft}, deformed in the infrared region $\phi\to-\infty$ by a $\mathbb{Z}_2$ twisted deformation. We showed that for $R_{AdS}\ge l_s$ $(k\ge 1)$, there is a similarly strong case for the hypothesis that an analogous deformed symmetric product provides an effective description of the dynamics of states whose scaling dimensions are well below the spacetime central charge.  

The situation is similar to the difference between asymptotically free QFT's, like four-dimensional non-abelian gauge theories, and theories like QED, which are inherently effective descriptions of something else. The case $R_{AdS}=l_s$ (or $k=1$) is analogous from this perspective to finite gauge theories, like $N=4$ SYM. In our case, it serves as a dividing point between $k<1$, where the long string coupling $g_\ell$ \eqref{gsl} goes to zero in the UV $(\phi\to\infty)$, and $k>1$, where it diverges there. For $k=1$, it is natural to conjecture that the spacetime CFT is that of \cite{Balthazar:2021xeh}, the symmetric product \eqref{symprodcft}, with a $\mathbb{Z}_2$ twisted deformation with the profile \eqref{profztwo}. This deformation was missing in \cite{Giribet:2018ada,Gaberdiel:2024dva}.

An interesting question, raised by the referee of the original version of this paper, is whether for $k=1$ one can find a holographic dual for the SW orbifold \eqref{longs}, \eqref{symprodcft}, with the $\mathbb{Z}_2$ twisted deformation turned off. This question is beyond the scope of our paper, which takes as a starting point a particular $AdS_3$ background of string theory, and derives the boundary theory from a worldsheet analysis. In the resulting boundary theory, the above deformation is turned on. To find a dual to the undeformed SW orbifold, one would have to start with a different $AdS_3$ background, which we currently do not possess.   

Note also that the twisted sector
deformation is not really a modulus of the boundary CFT that can be tuned, since one
can change its coupling by a shift of the radial coordinate $\phi$. Thus, from the boundary
CFT point of view, we have two distinct theories, in which this coupling is one and zero, respectively. The former appears in the $k = 1$ holographic duality that we studied, while the latter may or may not appear in some other duality, which is not continuously connected to our analysis.\footnote{The theories we considered in this paper have RR moduli, and the above discussion needs to be reexamined in their presence. We leave this to future work.} 

For $k>1$, we seem to arrive at the following picture. The deformed symmetric product, where we turn off the deformation \eqref{deformblock}, provides an effective description of the spacetime CFT at energies well below the black hole threshold. Adding this deformation back seems to signal the breakdown of the effective description due to the black holes, but the precise relation between the two remains to be understood.  

We illustrated our discussion with a few classes of backgrounds in section \ref{sec3}. It might be interesting to consider other backgrounds with or without supersymmetry. For example, one can study the pure $AdS_3$ background, where $\mathcal{N}$ in section \ref{sec2} is empty, and $k=4/7$ (see \eqref{cn}). In this case, the SW theory \eqref{longs} consists of two fields, $(\phi,\psi_\phi)$. In a sense, it is the minimal example of the correspondence, and may serve as an interesting toy model.

\section{Note added}

A few months after the appearance of this paper, it was argued in \cite{Eberhardt:2025sbi} that for $k=1$ ($k=3$ in the bosonic string), one can define a new string theory on $AdS_3$, that only includes the (spectral flowed) continuous representations, and is dual to a symmetric orbifold CFT. In this section, we describe the results of \cite{Eberhardt:2025sbi} from the perspective of \cite{Balthazar:2021xeh,Eberhardt:2021vsx} and this paper. We also comment on some related issues.
\medskip

\noindent
The questions that motivated this note are:
\begin{itemize}
\item How is the theory of \cite{Eberhardt:2025sbi} related to  standard string theory on $AdS_3$ with $k=1$, in particular string theory on $AdS_3\times \mathbb T^3$, which includes as a special case the theory with $k=1$ discussed in section \ref{sec32}?
\item How does the large chiral algebra of the symmetric orbifold CFT appear from the bulk $AdS_3$ point of view?
\item What can one learn from this construction about more general $AdS_3/CFT_2$ dualities?
\end{itemize}

As discussed in \cite{Balthazar:2021xeh,Eberhardt:2021vsx} and in this paper, for general $k$, perturbative string theory on $AdS_3\times\mathcal{N}$ gives rise to the spacetime CFT \eqref{symprodcft}, a symmetric product CFT whose seed is given by \eqref{longs}. Importantly, the symmetric product structure is modified by a $\mathbb{Z}_2$ twisted deformation, whose form was discussed in \cite{Balthazar:2021xeh,Eberhardt:2021vsx} and in sections \ref{sec2}, \ref{sec3}. For $k<1$, the case discussed in \cite{Balthazar:2021xeh}, this description can be extended to arbitrarily high energies, while for $k>1$ it only captures the perturbative string states, as discussed above. 

A natural question is what happens at the transition point between these regimes, $k=1$. This can be discussed from the perspective of the boundary CFT, or from that of the bulk string theory on $AdS_3$. We next describe the two perspectives on this question, starting with the former. 

A useful starting point is the theory with $k<1$ studied in \cite{Balthazar:2021xeh}, which, as mentioned above, is better defined mathematically than that with $k>1$. The dual CFT is in this case a symmetric product with seed $\mathbb{R}_\phi\times\mathcal{N}$, and a $\mathbb{Z}_2$ twisted wall that prevents the system from exploring the strong coupling region. One may expect that by focusing on physics far from the wall, \ie by restricting to large $\phi$ in \eqref{longs}, we get a theory in which the wall is absent. This can be done by formally setting the coefficient of the $\mathbb{Z}_2$ twisted deformation, $\mu$, to zero. 

For generic $k$, this attempt fails, due to the linear dilaton with slope $Q_\ell$ given in eq. \eqref{Qell}. This linear dilaton causes amplitudes of generic (real) momentum modes to be ill-defined. If one starts with the theory with the $\mathbb{Z}_2$ twisted wall present, and tries to remove it by sending its location to $\phi\to-\infty$, the limit is singular.\footnote{One can define bulk amplitudes, in the sense of \cite{Aharony:2004xn}, but those involve in general operators with imaginary momentum. Moreover, the sets of bulk amplitudes (that are characterized by the sum rules satisfied by the momenta) are different at different genera (orders of string perturbation theory), and have contributions due to LSZ poles \cite{Aharony:2004xn}.} This is related to the fact that for general $k$ the theory is not translationally invariant in $\phi$, even in the absence of the $\mathbb{Z}_2$ twisted wall.   

The only case in which this procedure is potentially sensible is $k=1$. In that case, $Q_\ell$ vanishes, and setting $\mu= 0$ one recovers translational invariance in the $\phi$ direction. In the resulting theory, one can study amplitudes that are proportional to the (infinite) length of $\phi$, like we do in QFT and string theory in flat spacetime. However, this theory is not continuously connected to that with $\mu\not=0$. In particular, in this theory, states with positive and negative $\phi$ momentum are taken to be independent. For any $\mu\not=0$, they are related by reflection from the wall. 

A perhaps useful analog is a non-compact orbifold CFT, \eg $\mathbb{R}^n/\mathbb{Z}_2$. This CFT has a continuum of delta-function normalizable states that propagate in the bulk of $\mathbb{R}^n$, and normalizable states localized near the $\mathbb{Z}_2$ fixed point. There are three different theories that one can discuss in this context: (1) the CFT on $\mathbb{R}^n$; (2) the untwisted sector of the CFT on $\mathbb{R}^n/\mathbb{Z}_2$; (3) the full CFT on $\mathbb{R}^n/\mathbb{Z}_2$. Theories (1) and (2) are closely related but are distinct. In particular, while theory (1) is modular invariant, (2) is not, and it requires the twisted sector to become (3), which is modular invariant. And, of course, the CFT (1) is disconnected from (2), (3), in the sense that for general $n$ there is no marginal deformation that connects them. Physically, theories (1) and (2) differ in the absence of the  $\mathbb{Z}_2$ identification in the former. Dropping this identification makes sense if one is studying local scattering processes that occur infinitely far from the fixed point. 

The boundary CFT of string theory on $AdS_3\times\mathcal{N}$ with $k=1$ is the analog of theory (3) above. On the other hand, the theory studied in \cite{Eberhardt:2025sbi} is the analog of theory (1). The restriction of the full string theory on $AdS_3\times\mathcal{N}$ to the theory of continuous series states with $|w|\ge 1$ is the analog of theory (2). In this sense, the theory of \cite{Eberhardt:2025sbi} is disconnected from standard string theory on $AdS_3\times\mathcal{N}$, just like in the case of CFT's (1) and (3).

The discussion above was from the point of view of the boundary theory, the CFT dual to string theory on $AdS_3\times\mathcal{N}$. It is natural to ask how it looks from the bulk point of view. An interesting question is the following. The symmetric product without the $\mathbb{Z}_2$ twisted wall has a much larger chiral algebra than the full string theory on $AdS_3$. How does one see this symmetry from the bulk perspective? We next answer this question. 

For concreteness, we discuss a special case -- the operator $\partial_x\phi$ that is not holomorphic in the full string theory on $AdS_3\times\mathcal{N}$, but becomes holomorphic in its large $\phi$ limit. The mechanism in operation for this operator is the same for other holomorphic operators in the symmetric product theory. 

The breaking of holomorphy for the operator $\partial_x\phi$ is discussed in section 7.2 of \cite{Balthazar:2021xeh}. It is noted there (see also \eqref{ddphimap} here) that the operator $\partial_{\bar{x}}\partial_x\phi$ corresponds in the bulk to an operator with $j=1-\frac{k}2$, $m=\bar m=\frac{k}2$. For $k<1$, this operator is non-normalizable, but it sits on an LSZ pole (in the sense of \cite{Aharony:2004xn}). By deforming away from the pole (see eq. (7.2) in \cite{Balthazar:2021xeh}) and carefully taking the limit, it was shown in \cite{Balthazar:2021xeh} that this operator does not vanish but gives a normalizable operator, (7.5), (7.10), which corresponds in the boundary theory to the $\mathbb{Z}_2$ twisted deformation.

For $k=1$, the analysis of \cite{Balthazar:2021xeh} has an interesting twist. In that case, the bulk operator corresponding to  $\partial_{\bar{x}}\partial_x\phi$ has $j=\frac12$, and thus lies at the boundary between the delta-function normalizable and non-normalizable branches. Therefore, one can deform it as in (7.2) in two different ways. One is as done there, which takes this operator into the non-normalizable branch, leading us to study operators of the form $e^{\beta\phi}\partial_x\phi $ with real $\beta>0$.\footnote{$\beta=\epsilon Q$ in the notation of eq. (7.2). Note that this involves correcting a typo in \cite{Balthazar:2021xeh}, where in the line above (7.2) $Q$ is replaced by $Q_\ell$.} The other, corresponding to imaginary $\epsilon=ip$, takes it into the delta-function normalizable branch,\footnote{Note that for the consistency of this procedure, it is important that the coefficient of $\epsilon$ in the expression for $m$ in eq. (7.2) of \cite{Balthazar:2021xeh} vanishes for $k=1$. The full expression for $m$ has an extra term, that is quadratic in $\epsilon$, but that is not a problem since it remains real when $\epsilon$ is imaginary.} and to operators of the above form with $\beta=\sqrt2ip$. 

Approaching $\epsilon\to 0$ from real values leads to the same conclusions as in \cite{Balthazar:2021xeh}:  $\partial_{\bar{x}}\partial_x\phi$ does not vanish, and is given by the $\mathbb{Z}_2$ twisted operator that deforms the symmetric orbifold. On the other hand, if we approach $\epsilon=0$ from imaginary values, we can avoid the LSZ pole, and view the operator $\partial_{\bar{x}}\partial_x\phi$ as a limit of delta-function normalizable operators, for which one can use the construction of \cite{Eberhardt:2025sbi}. However, this comes with a price. If we take $\epsilon=ip$, in the limit $p\to 0$ we are dealing with real $\phi$ momentum that goes to zero. To avoid the LSZ pole, we have to take $\phi\to\infty$ as $p\to 0$. This recovers the conclusion we got from the boundary CFT point of view -- that the decoupled theory of the continuous series states in $AdS_3$ lives at infinite $\phi$. And, it shows that in the theory describing the infinite $\phi$ region, the operator $\partial_x\phi$ is holomorphic, as one would expect from the boundary perspective.

It is useful to contrast the above discussion with what happens for $k\not=1$. The construction of \cite{Balthazar:2021xeh} allows one to study delta-function normalizable operators in the boundary CFT, of the form $e^{\beta\phi}\partial_x\phi $, with $\beta=-\frac{Q_\ell}2+ip$ (eq. (4.7) in \cite{Balthazar:2021xeh}). However, in order to reach the operator $\partial_x\phi$ of interest here, one has to analytically continue to a finite imaginary value of $p$ that corresponds to $\beta=0$. The resulting operator is non-normalizable, and therefore, it does not satisfy the decoupling discussed above.

As mentioned earlier, while the explicit discussion above was for the operator $\partial_x\phi$, it is easy to generalize it in two directions. One is to the infinite set of conserved currents that, like $\partial_x\phi$, have the property that they are conserved in the $\phi\to\infty$ theory, but not in the full string theory on $AdS_3\times\mathcal{N}$ with $k=1$. Another is to currents that are conserved in the full theory. Such currents are usually described by operators in the winding zero sector \cite{Giveon:1998ns,Kutasov:1999xu}, but one can construct them in the $w=-1$ sector as well. The key point is that, like in the discussion above, for $k=1$ they have $j=\frac12$, and can be viewed as limits from the continuum. We will not describe the details here.

Coming back to the questions raised at the beginning of this section, we have already addressed the first two. We saw that the theory described in \cite{Eberhardt:2025sbi} is related to the one obtained from the full string theory on $AdS_3$ with $k=1$ by taking the limit $\phi\to\infty$, and focusing on the local physics of delta-function normalizable operators. We also explained how the large chiral algebra of the boundary CFT, the symmetric product with seed $\mathbb{R}\times \mathcal{N}$, is obtained in the full $AdS_3$ theory, by studying delta-function normalizable operators in the limit where the $\phi$ momentum goes to zero. 

We also addressed the third question. The standard construction of string theory on $AdS_3$, as near-horizon geometries of systems of intersecting NS fivebranes and fundamental strings, leads to backgrounds that contain both discrete (normalizable) and continuous (delta-function normalizable) states, which interact with each other. This is the general situation in $AdS_3/CFT_2$, for any $k$, including $k=1$. In the construction of \cite{Eberhardt:2025sbi}, one discards the normalizable states and drops the identification between delta-function normalizable states with positive and negative radial momenta. In that sense, it is disconnected from the standard construction, just like CFT on $\mathbb{R}^n$ is disconnected from that on $\mathbb{R}^n/\mathbb{Z}_2$. It remains to be seen how much it can teach us about string theory on $AdS_3$.

\bigskip

\section*{Acknowledgements} 

We thank O. Aharony, A. Dei and E. Martinec for discussions. The work of SC received funding from the Department of Physics at The Ohio State University. 
The work of AG was supported in part by the ISF (grant number 256/22). The work of DK was supported in part by DOE grant DE-SC0009924 and the FACCTS Program at the University of Chicago.



\begin{thebibliography}{10}

\bibitem{Balthazar:2021xeh}
B.~Balthazar, A.~Giveon, D.~Kutasov and E.~J. Martinec, \emph{{Asymptotically
  free AdS$_{3}$/CFT$_{2}$}},
  \href{https://doi.org/10.1007/JHEP01(2022)008}{\emph{JHEP} {\bfseries 01}
  (2022) 008} [\href{https://arxiv.org/abs/2109.00065}{{\ttfamily
  2109.00065}}].

\bibitem{Eberhardt:2021vsx}
L.~Eberhardt, \emph{{A perturbative CFT dual for pure NS\textendash{}NS
  AdS$_{3}$ strings}}, \href{https://doi.org/10.1088/1751-8121/ac47b2}{\emph{J.
  Phys. A} {\bfseries 55} (2022) 064001}
  [\href{https://arxiv.org/abs/2110.07535}{{\ttfamily 2110.07535}}].

\bibitem{Giveon:2005mi}
A.~Giveon, D.~Kutasov, E.~Rabinovici and A.~Sever, \emph{{Phases of quantum
  gravity in AdS(3) and linear dilaton backgrounds}},
  \href{https://doi.org/10.1016/j.nuclphysb.2005.04.015}{\emph{Nucl. Phys. B}
  {\bfseries 719} (2005) 3}
  [\href{https://arxiv.org/abs/hep-th/0503121}{{\ttfamily hep-th/0503121}}].

\bibitem{Seiberg:1999xz}
N.~Seiberg and E.~Witten, \emph{{The D1 / D5 system and singular CFT}},
  \href{https://doi.org/10.1088/1126-6708/1999/04/017}{\emph{JHEP} {\bfseries
  04} (1999) 017} [\href{https://arxiv.org/abs/hep-th/9903224}{{\ttfamily
  hep-th/9903224}}].

\bibitem{Giribet:2018ada}
G.~Giribet, C.~Hull, M.~Kleban, M.~Porrati and E.~Rabinovici,
  \emph{{Superstrings on AdS$_{3}$ at $k =$ 1}},
  \href{https://doi.org/10.1007/JHEP08(2018)204}{\emph{JHEP} {\bfseries 08}
  (2018) 204} [\href{https://arxiv.org/abs/1803.04420}{{\ttfamily
  1803.04420}}].

\bibitem{Giribet:2020mkc}
G.~Giribet, \emph{{String theory on AdS$_3\times {M}_7$ in the tensionless
  limit}}, \href{https://doi.org/10.1142/S0218271820300050}{\emph{Int. J. Mod.
  Phys. D} {\bfseries 29} (2020) 2030005}
  [\href{https://arxiv.org/abs/2003.02868}{{\ttfamily 2003.02868}}].

\bibitem{Gaberdiel:2024dva}
M.~R. Gaberdiel and V.~Sriprachyakul, \emph{{Tensionless strings on $AdS_3
  \times S^3 \times S^3 \times S^1$}},
  \href{https://arxiv.org/abs/2411.16848}{{\ttfamily 2411.16848}}.

\bibitem{Aharony:2004xn}
O.~Aharony, A.~Giveon and D.~Kutasov, \emph{{LSZ in LST}},
  \href{https://doi.org/10.1016/j.nuclphysb.2004.05.015}{\emph{Nucl. Phys. B}
  {\bfseries 691} (2004) 3}
  [\href{https://arxiv.org/abs/hep-th/0404016}{{\ttfamily hep-th/0404016}}].

\bibitem{Klemm:1990df}
A.~Klemm and M.~G. Schmidt, \emph{{Orbifolds by Cyclic Permutations of Tensor
  Product Conformal Field Theories}},
  \href{https://doi.org/10.1016/0370-2693(90)90164-2}{\emph{Phys. Lett. B}
  {\bfseries 245} (1990) 53}.

\bibitem{Argurio:2000tb}
R.~Argurio, A.~Giveon and A.~Shomer, \emph{{Superstrings on AdS(3) and
  symmetric products}},
  \href{https://doi.org/10.1088/1126-6708/2000/12/003}{\emph{JHEP} {\bfseries
  12} (2000) 003} [\href{https://arxiv.org/abs/hep-th/0009242}{{\ttfamily
  hep-th/0009242}}].

\bibitem{Elitzur:1998mm}
S.~Elitzur, O.~Feinerman, A.~Giveon and D.~Tsabar, \emph{{String theory on
  $AdS_3 \times S^3 \times S^3 \times S^1$}},
  \href{https://doi.org/10.1016/S0370-2693(99)00101-X}{\emph{Phys. Lett. B}
  {\bfseries 449} (1999) 180}
  [\href{https://arxiv.org/abs/hep-th/9811245}{{\ttfamily hep-th/9811245}}].

\bibitem{deBoer:1999gea}
J.~de~Boer, A.~Pasquinucci and K.~Skenderis, \emph{{AdS / CFT dualities
  involving large 2-D N=4 superconformal symmetry}},
  \href{https://doi.org/10.4310/ATMP.1999.v3.n3.a5}{\emph{Adv. Theor. Math.
  Phys.} {\bfseries 3} (1999) 577}
  [\href{https://arxiv.org/abs/hep-th/9904073}{{\ttfamily hep-th/9904073}}].

\bibitem{Gukov:2004ym}
S.~Gukov, E.~Martinec, G.~W. Moore and A.~Strominger, \emph{{The Search for a
  holographic dual to $AdS_3 \times S^3 \times S^3 \times S^1$}},
  \href{https://doi.org/10.4310/ATMP.2005.v9.n3.a3}{\emph{Adv. Theor. Math.
  Phys.} {\bfseries 9} (2005) 435}
  [\href{https://arxiv.org/abs/hep-th/0403090}{{\ttfamily hep-th/0403090}}].

\bibitem{Tong:2014yna}
D.~Tong, \emph{{The holographic dual of $AdS_{3} \times S^{3} \times S^{3}
  \times S^{1}$}}, \href{https://doi.org/10.1007/JHEP04(2014)193}{\emph{JHEP}
  {\bfseries 04} (2014) 193} [\href{https://arxiv.org/abs/1402.5135}{{\ttfamily
  1402.5135}}].

\bibitem{Eberhardt:2017fsi}
L.~Eberhardt, M.~R. Gaberdiel, R.~Gopakumar and W.~Li, \emph{{BPS spectrum on
  AdS$_3\times $S$^3 \times $S$^3 \times $S$^1$}},
  \href{https://doi.org/10.1007/JHEP03(2017)124}{\emph{JHEP} {\bfseries 03}
  (2017) 124} [\href{https://arxiv.org/abs/1701.03552}{{\ttfamily
  1701.03552}}].

\bibitem{Eberhardt:2017pty}
L.~Eberhardt, M.~R. Gaberdiel and W.~Li, \emph{{A holographic dual for string
  theory on
  AdS$_{3}$\texttimes{}S$^{3}$\texttimes{}S$^{3}$\texttimes{}S$^{1}$}},
  \href{https://doi.org/10.1007/JHEP08(2017)111}{\emph{JHEP} {\bfseries 08}
  (2017) 111} [\href{https://arxiv.org/abs/1707.02705}{{\ttfamily
  1707.02705}}].

\bibitem{Witten:2024yod}
E.~Witten, \emph{{Instantons and the Large N=4 Algebra}},
  \href{https://arxiv.org/abs/2407.20964}{{\ttfamily 2407.20964}}.

\bibitem{Giveon:1999jg}
A.~Giveon and M.~Rocek, \emph{{Supersymmetric string vacua on AdS(3) x N}},
  \href{https://doi.org/10.1088/1126-6708/1999/04/019}{\emph{JHEP} {\bfseries
  04} (1999) 019} [\href{https://arxiv.org/abs/hep-th/9904024}{{\ttfamily
  hep-th/9904024}}].

\bibitem{Berenstein:1999gj}
D.~Berenstein and R.~G. Leigh, \emph{{Space-time supersymmetry in AdS(3)
  backgrounds}},
  \href{https://doi.org/10.1016/S0370-2693(99)00623-1}{\emph{Phys. Lett. B}
  {\bfseries 458} (1999) 297}
  [\href{https://arxiv.org/abs/hep-th/9904040}{{\ttfamily hep-th/9904040}}].

\bibitem{Giveon:2003ku}
A.~Giveon and A.~Pakman, \emph{{More on superstrings in AdS(3) x N}},
  \href{https://doi.org/10.1088/1126-6708/2003/03/056}{\emph{JHEP} {\bfseries
  03} (2003) 056} [\href{https://arxiv.org/abs/hep-th/0302217}{{\ttfamily
  hep-th/0302217}}].

\bibitem{Eberhardt:2025sbi}
L.~Eberhardt and M.~R. Gaberdiel, \emph{{A localising AdS$_3$ sigma model}},
  \href{https://arxiv.org/abs/2505.09226}{{\ttfamily 2505.09226}}.

\bibitem{Giveon:1998ns}
A.~Giveon, D.~Kutasov and N.~Seiberg, \emph{{Comments on string theory on
  AdS(3)}}, \href{https://doi.org/10.4310/ATMP.1998.v2.n4.a3}{\emph{Adv. Theor.
  Math. Phys.} {\bfseries 2} (1998) 733}
  [\href{https://arxiv.org/abs/hep-th/9806194}{{\ttfamily hep-th/9806194}}].

\bibitem{Kutasov:1999xu}
D.~Kutasov and N.~Seiberg, \emph{{More comments on string theory on AdS(3)}},
  \href{https://doi.org/10.1088/1126-6708/1999/04/008}{\emph{JHEP} {\bfseries
  04} (1999) 008} [\href{https://arxiv.org/abs/hep-th/9903219}{{\ttfamily
  hep-th/9903219}}].

\end{thebibliography}

\providecommand{\href}[2]{#2}\begingroup\raggedright\endgroup

\end{document}